\documentclass[twocolumn,twoside,slac_two]{revtex4}
\usepackage{graphicx}
\usepackage{fancyhdr}
\begin{document}

\title{A double component in the prompt emission of GRB 090618}


\author{L. Izzo$^{1}$, R. Ruffini$^{1,2}$, A. V. Penacchioni$^{1,2}$, C. L. Bianco$^{1}$, M. Muccino$^{1}$, L. Caito$^{1}$, B. Patricelli$^{1}$}
\affiliation{$^{1}$University of Rome Sapienza and ICRANet, Dip. Fisica, p.le A. Moro 2, 00185 - Rome, Italy;\\
$^{2}$ Universite de Nice Sophia Antipolis, CEDEX 2, Grand Chateau Parc Valrose, Nice, France.\\}

\author{S. K. Chakrabarti$^{3,4}$, A. Nandi$^{4}$}
\affiliation{$^{3}$S. N. Bose National Center for Basic Sciences, Salt Lake, Kolkata - 700098, India;\\
$^{4}$Indian Center for Space Physics, Garia, Kolkata - 700084, India.}
\begin{abstract}
GRB 090618 offered an unprecedented opportunity to have coordinated data, by the
best of the X and Gamma Ray observatories, of the nearest (z = 0.54) energetic
source ($10^{54}$ erg). Using the Fermi-GBM observations of this GRB, we have analyzed this source to explore 
the possibility of having components yet to be observed in other sources. 
We show that it is not possible to interpret GRB 090618 within the framework 
of the traditional single component GRB model. We argue that the observation 
of the first episode of duration of around 50s could not be a part of a 
canonical GRB, while the residual emission could be modeled easily with 
the models existing in literature. In this work we have considered 
the case of the fireshell scenario.
\end{abstract}

\maketitle

\thispagestyle{fancy}


\section{INTRODUCTION}

We have studied the emission in the fireshell model of GRB 090618, which is one of the closest ($z=0.54$)
and the most energetic ($E_{iso} = 2.4 \times 10^{54}$ erg) GRBs. It has been observed by many satellites,
namely Fermi, Swift, Konus-WIND, AGILE, RT-2 and Suzaku. We have analyzed the emission of this
GRB first identifying the transparency emission, the P-GRB, and then using different spectral models 
with XSPEC. The fundamental parameters to be determined in the fireshell model, \cite{1}, in order 
to obtain information on the spectral energy emission, 
are the dyadosphere energy, the baryon loading and the density and porosity of the CBM. We
found that in this GRB there exists two different components. The first component lasts 50 s 
with a spectrum showing a very clear thermal component evolving, between $kT = 60$ keV and 
$kT = 14$ keV, and a radius increasing between 9000 km and 50000 km, with an estimate mass of $\sim$ 10 M$\odot$. 
The second component is a canonical long GRB with a Lorentz gamma factor at the transparency 
of $\Gamma$ = 490, a temperature at transparency of 25.48 keV and with a characteristic size 
of the CBM cloud of $R_{cl} \approx 10^{15}$ cm which generated the observed luminosity. We 
confirm that the second episode corresponds to a canonical GRB, while the first episode do not. 
Indeed, it appears to be related to the progenitor of the collapsing bare core -- defined by us as
the ''proto black hole'' (Izzo et al. A$\&$A submitted) -- leading to the black hole formation.

\subsection{A BRIEF DESCRIPTION OF THE FIRESHELL SCENARIO}

Within the fireshell scenario, all GRBs originate from an optically thick $e^+e^-$ plasma in thermal 
equilibrium, as a result of a gravitational collapse to form a black hole. This plasma has a total 
energy of $E_{tot}^{e^+e^-}$. The annihilation of $e^+e^-$ pairs occurs gradually and is confined 
in an expanding shell called ''fireshell''. This plasma engulfs the baryonic material (of mass 
$M_{B}$) left over in the process of gravitational collapse. The baryon loading is measured by 
the dimensionless parameter $B = M_B c^2/E_{tot}^{e^+e^-}$. The fireshell self-accelerates to 
ultra-relativistic velocities until it reaches the transparency, when all the photons are emitted
 in what is called the P-GRB. The remaining accelerated baryonic matter starts then to slow down 
due to the collisions with the Circum Burst Medium (CBM), of 
average density $n_{CBM}$. This collision between baryons and the CBM will give rise to the 
extended afterglow emission, which represents the residual high-energy emission observed in GRBs.

\section{DATA ANALYSIS}

\subsection{Data reduction and light curve}

\begin{figure*}[t]
\begin{center}
\begin{tabular}{|c|c|}
\hline
\includegraphics[height=5cm,width=8.5cm,angle=0]{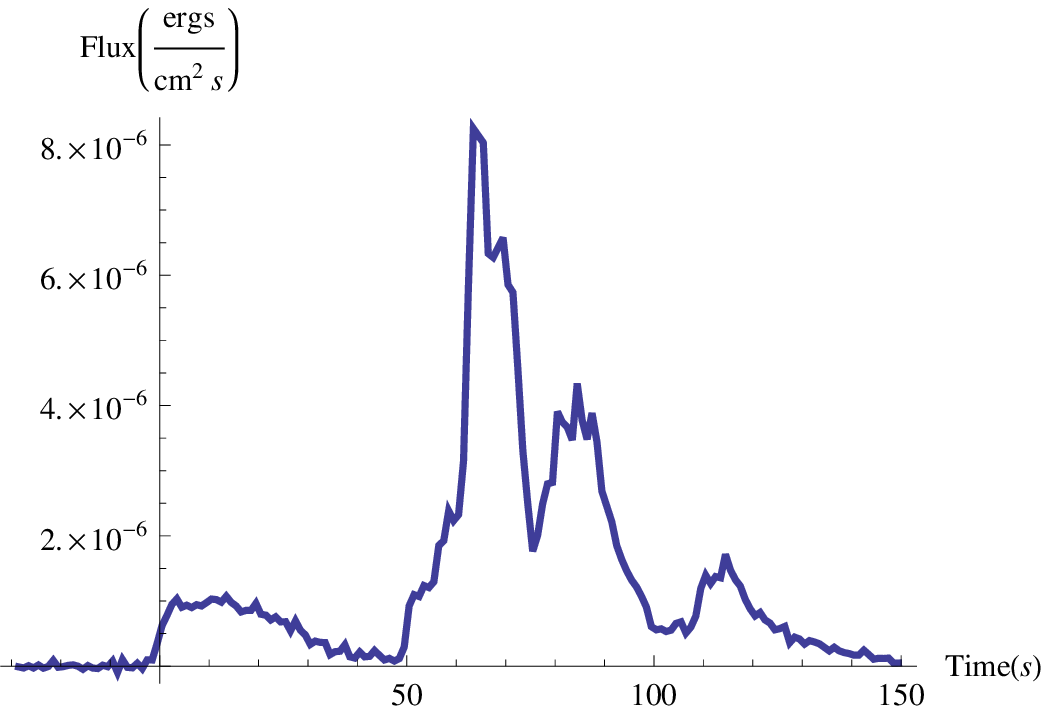} &
\includegraphics[height=5cm,width=8.5cm,angle=0]{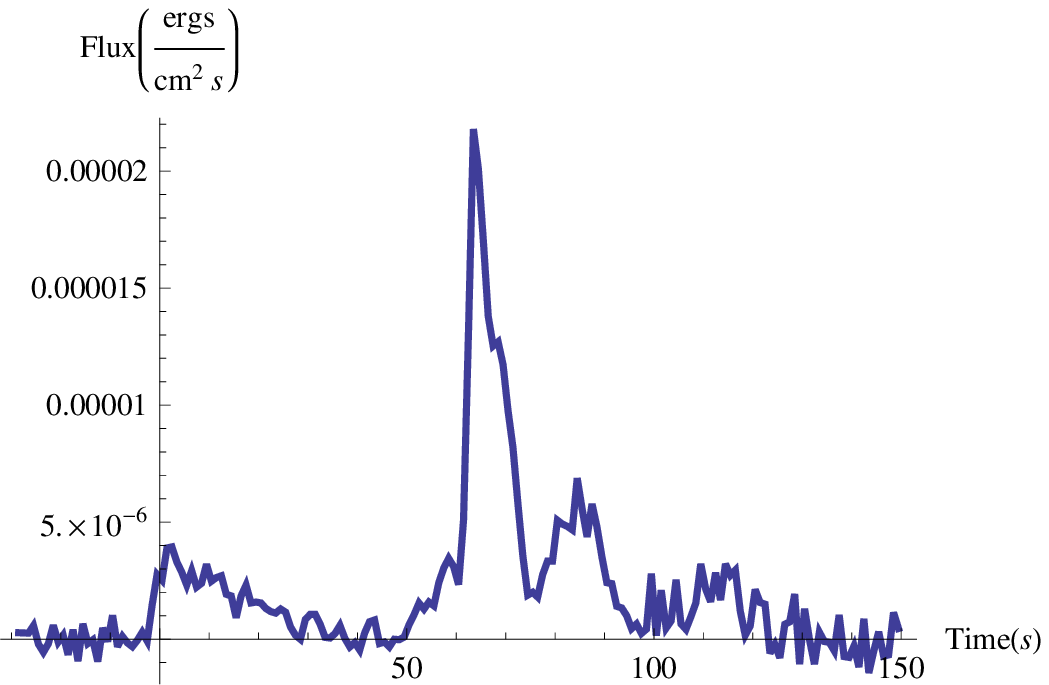}\tabularnewline
\hline
\end{tabular}
\caption{Fermi-GBM flux light curve of GRB 090618 referring to the NaI (8-440 keV, \emph{left panel}) 
and BGO (260 keV - 40 MeV, \emph{right panel}) detectors.}
\label{fig:1}
\end{center}\end{figure*}

We made use of Swift-BAT and XRT \cite{2} data, together with the Fermi-GBM \cite{3} and Coronas
PHOTON-RT2 \cite{4} ones. The data reduction was done using the Heasoft packages \cite{5} 
for BAT and XRT, plus the Fermi-Science tools for GBM. The Swift-BAT light curve was obtained in 
the band (15- 150 keV) using the standard procedure. The Fermi-GBM light curves are shown 
in Fig. \ref{fig:1}.

\subsection{Spectral analysis: identification of the P-GRB}

The identification of the P-GRB emission is fundamental since it allows us
to determine the value of the Baryon Loading and other physical properties 
of the fireshell plasma at the initial stage, such as the temperature and the Lorentz gamma 
factor at the transparency and the lab radius when the P-GRB 
emission would happen. From the observations we obtain the
the P-GRB energy and the temperature of the black body due to the P-GRB emission \cite{6}.
For this reason we proceeded with the time resolved spectral analysis
 of GRB 090618 dividing the emission in several time intervals. 
We then made a data fitting procedure with XSPEC [3] using two spectral 
models: a Band function and black body plus a power-law component 
(bb+po). The results are shown in table \ref{tab:1}.

\begin{table*}
\begin{center}
\caption{Time-resolved spectral analysis of GRB 090618.} % title ofable
\label{tab:1} % is used to refer this table in the text
\begin{tabular}{l c c c c c c c}
\hline\hline
Time Interval  & $\alpha$ & $\beta$ & $E_0^{BAND}$ (keV) & conv. factor (ergs/cm$^2$/ctg) & $\tilde{\chi}^2$ & $\gamma$ & $E_0^{COP}$ (keV)\\ % table heading
\hline % inserts single horizontal line 
0 - 50 & -0.86 $\pm$ 0.10 & -2.30 $\pm$ 0.10 & 152.4 $\pm$ 37.4 & 1.50 $\times$ 10$^{-9}$ & 1.16 &0.79 $\pm$ 0.07 & 140.7 $\pm$ 17.9 \\
        & -1.48 $\pm$ 0.10 & -2.86 $\pm$ 0.86 & 227.6 $\pm$ 50 & 9.03 $\times$ 10$^{-9}$ & 1.06 & 2.20 $\pm$ 1.87 & 499.9 $\pm$ 1260.1 \\  
50 - 57 & -1.10 $\pm$ 0.10 & -3.23 $\pm$ 6.33 & 174.0 $\pm$ 51.8 & 1.32 $\times$ 10$^{-9}$ & 1.28 & 1.04 $\pm$ 0.07& 169.1 $\pm$ 28.3 \\
        & -3.03 $\pm$ 1.57 & -2.30 $\pm$ 0.10 & 701.7 $\pm$ 50 & 5.94 $\times$ 10$^{-9}$ & 1.11 & 1.47 $\pm$ 2.96& 255.5 $\pm$ 540.0 \\
57 - 68 & -0.84 $\pm$ 0.03 & -2.44 $\pm$ 0.37 & 198.5 $\pm$ 17.0 & 1.64 $\times$ 10$^{-9}$ & 1.81 & 0.84 $\pm$ 0.02& 216.7 $\pm$ 10.7 \\
        & -1.34 $\pm$ 0.40 & -3.78 $\pm$ 1.21 & 565.4 $\pm$ 132.16 & 7.22 $\times$ 10$^{-9}$ & 1.11 & 1.3 $\pm$ 0.21& 500.0 $\pm$ 233.0 \\
68 - 76 & -1.03 $\pm$ 0.03 & -2.30 $\pm$ 0.10 & 175.7 $\pm$ 16.6 & 1.41 $\times$ 10$^{-9}$ & 1.62  & 0.98 $\pm$ 0.02& 169.3 $\pm$ 8.11 \\
        & -1.75 $\pm$ 0.71 & -4.50 $\pm$ 6.99 & 548.9 $\pm$ 415.5 & 6.67 $\times$ 10$^{-9}$ & 0.96 & 2.22 $\pm$ 0.48& 500.0 $\pm$ 638.8 \\
76 - 103 & -1.06 $\pm$ 0.03 & -2.90 $\pm$ 0.28 & 124.4 $\pm$ 7.67 & 1.23 $\times$ 10$^{-9}$ & 1.27 & 1.03 $\pm$ 0.02& 128.7 $\pm$ 5.7 \\          
         & -2.25 $\pm$ 1.77 & -2.30 $\pm$ 0.10 & 520.1 $\pm$ 50 & 6.97 $\times$ 10$^{-9}$ & 0.88 & 2.076 $\pm$ 0.89 & 389.4 $\pm$ 365.4 \\          
103 - 150 & -2.56 $\pm$ 0.10 & -2.03 $\pm$ 0.02 & $<$ 100 & 1.17 $\times$ 10$^{-9}$ & 2.24  & 1.42 $\pm$ 0.06 & 93.3 $\pm$ 11.98 \\
          & -0.84 $\pm$ 1.00 & -0.57 $\pm$ 1.39 & $<$ 100 & 1.79 $\times$ 10$^{-8}$ & 1.07 & 2.076 $\pm$ 0.89 & 389.4 $\pm$ 365.438 \\
\hline
\end{tabular}
\end{center}
\end{table*}

\section{RESULTS}

\subsection{From 0 to 50s}

The results of this analysis showed the presence of a possible single black body component 
in the first two intervals considered in the Table \ref{tab:1}. The former one 
corresponds to a long precursor, lasting $\sim$ 50 s, which is characterized by 
a single fast rise and exponential decay (FRED) pulse.  

For this reason, we divided the emission into two main episodes --  
the first one lasting from 0 to 50s, and the second one from 50 to 150s. 
The first episode is well-fitted by a BB+po model of temperature 
$ kT = 29.84 \pm 1.38$ keV (which in principle is a distinctive 
feature of a P-GRB) and a photon index $\gamma = -1.67 \pm 0.03$. 
We know that the isotropic energy of the GRB is $E_{iso} = 
2.8 \times 10^{53}$ erg, while the energy emitted by the sole 
black body component in this first 50 s emission is $E_{BB,1st} = 
8.88 \times 10^{51}$ erg (the 3.2\% of the total energy emitted). 
This implies a baryon loading $B = 10^{-4}$. From the fireshell 
equations of motion we derived a theoretically predicted temperature 
at the transparency which, when corrected for the cosmological redshift of 
the source, gives $kT = 425$ keV. This value is in clear disagreement 
with the observed temperature which led us to conclude that the first 
episode cannot be considered as the transparency emission, or the P-GRB. 
Another peculiar feature is that the duration of this episode is much 
longer than the typical one considered for P-GRBs, which is at the most 
10s, see e.g. \cite{1}.

\subsection{The second episode as an independent GRB}

\begin{figure}
\includegraphics[height=8cm,width=6cm,angle=270]{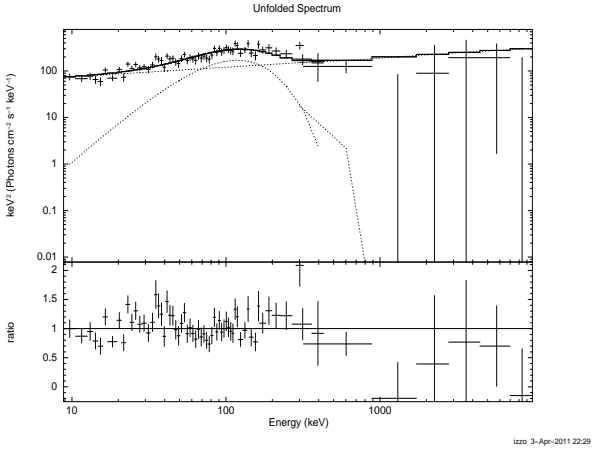}\\
\caption{Time-integrated spectra of the P-GRB of the second episode (from 50 to 54 s after the trigger time) of GRB 090618 fitted with the blackbody + power-law model, $\chi^2$ = 1.52.}\label{fig:2}
\end{figure}

\begin{figure*}[t]
\begin{center}
\begin{tabular}{|c|c|}
\hline
\includegraphics[height=8cm,width=5cm,angle=270]{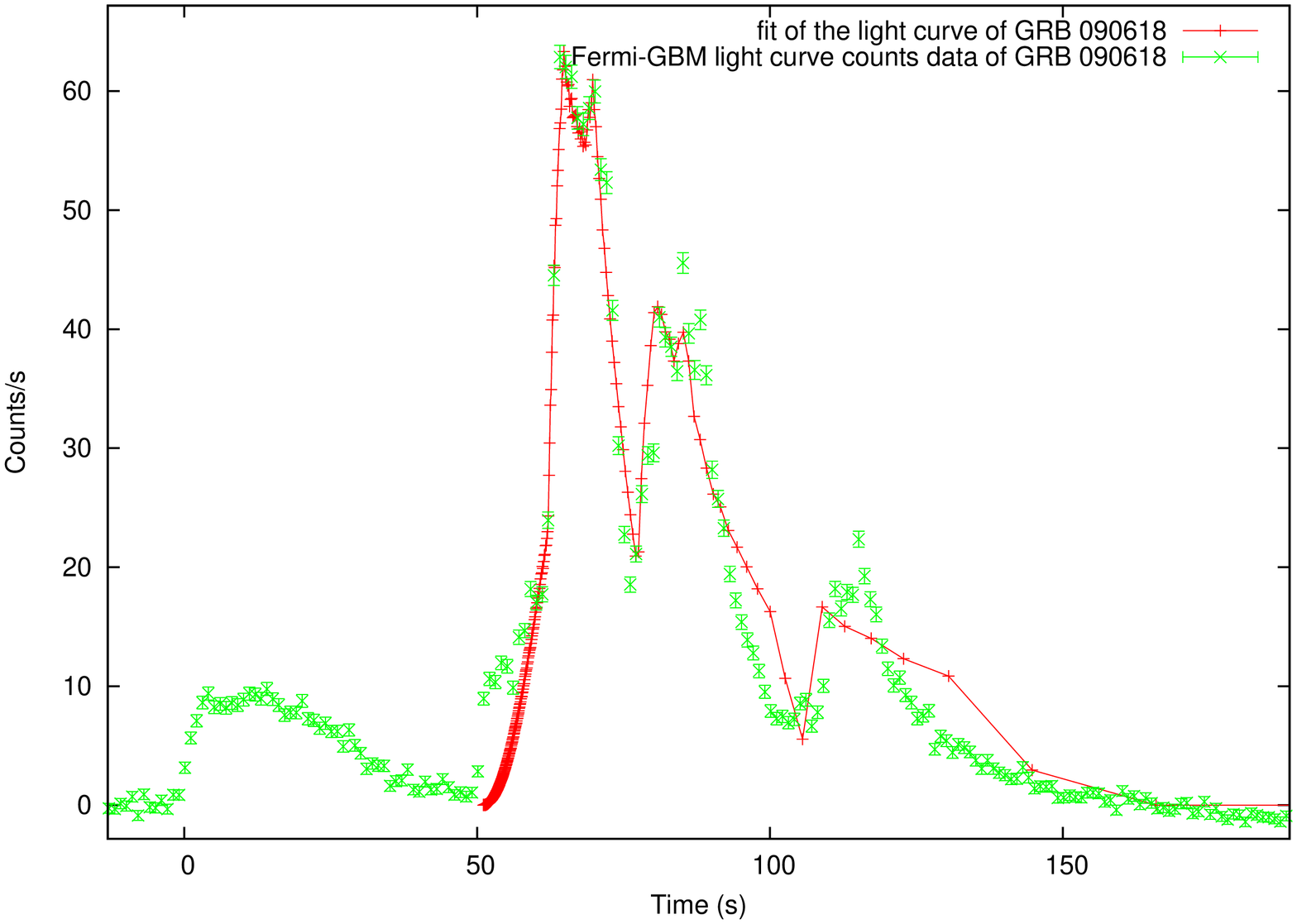} &
\includegraphics[height=8cm,width=5cm,angle=270]{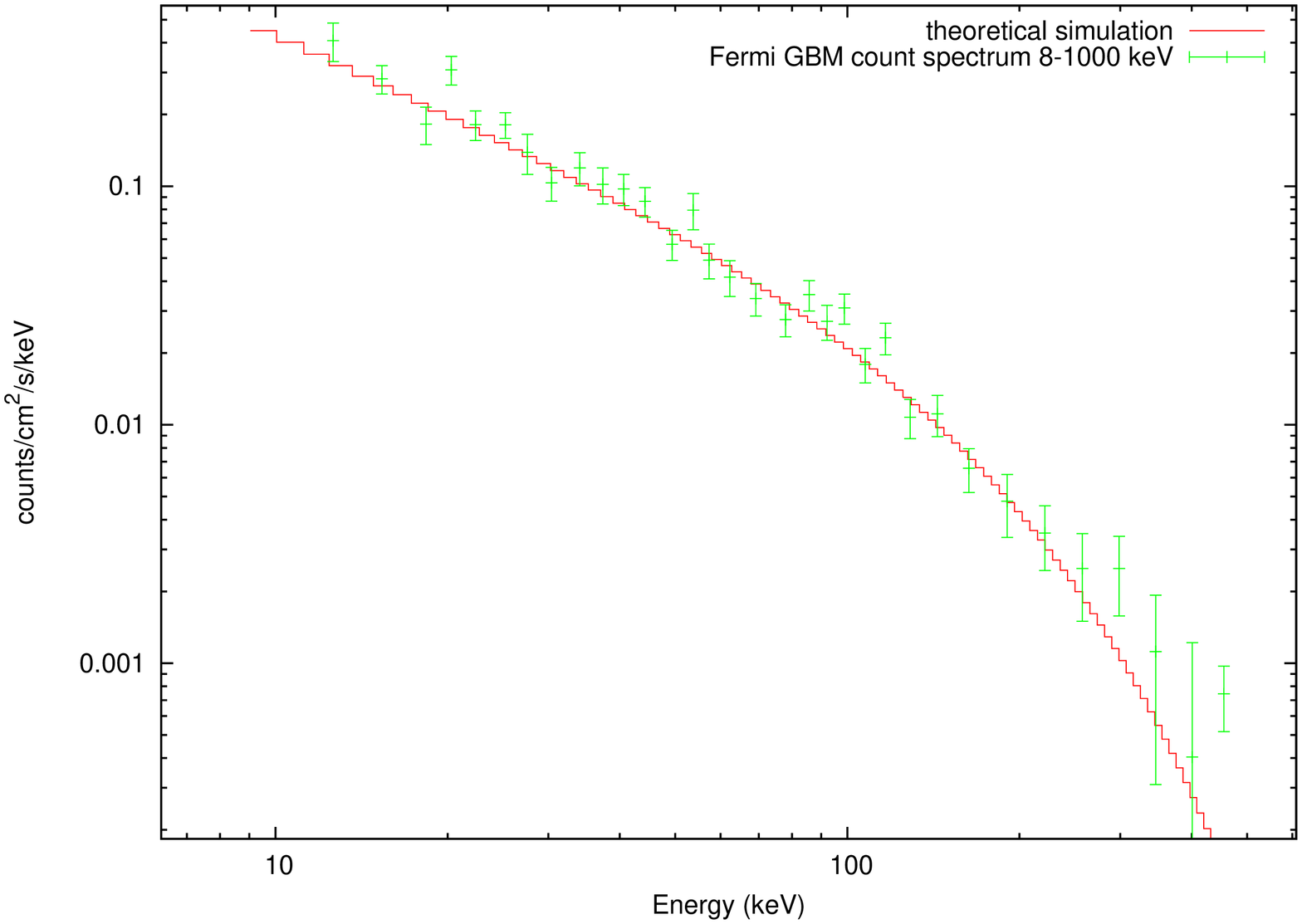}\tabularnewline
\hline
\end{tabular}
\caption{Simulated light curve (left) and time integrated (t0+58, t0+150 s) spectrum (8-440 keV, on the right) of the extended afterglow of GRB 090618.}
\label{fig:3}
\end{center}\end{figure*}

We tried to identify the P-GRB within the thermal emission observed 
in the interval B, so from 50s to 59s after the GBM trigger time. A detailed 
analysis allowed us to consider the first 4 s as due to the P-GRB emission. The P-GRB spectrum is well-fitted 
by a black body plus a power-law extra component, see Fig. \ref{fig:2} with temperature 
$kT = 25.48 \pm 2.04$ keV and a photon index $\gamma = 1.85 \pm 0.06$. The integrated 
spectrum of the remaining part is best fitted with a Band model (see Table \ref{tab:1}). 
The isotropic energy of the second episode is $E_{iso,2nd} = 2.37 \times 10^{53}$ erg. 
If we assume the equality $E_{iso} = E_{tot}^{e^+e^-}$ and a baryon loading $B = 
2 \times 10^{-3}$, we find that the P-GRB energy should correspond to the 2\% of 
the total energy of the GRB. Since from the fireshell equations we obtain a Lorentz 
gamma factor at the transparency of $\Gamma_0 = 490$, the corresponding theoretical, 
and cosmologically corrected, temperature at transparency should be $kT = 24.48$ keV, 
in agreement with the observations. So as a final result, this second episode can be 
considered as a canonical GRB in the fireshell scenario.
We have simulated in this model the GRB 090618 light curve and spectrum, that are 
shown in Fig.\ref{fig:3}.

\subsection{The first 50 s emission is not a GRB}

We then have checked if the first 50s emission could be considered to be an independent GRB. We attempted a first 
interpretation by assuming the first 6 s as the P-GRB component, as opposed to the remaining 44 s 
as the possible extended afterglow. A possible combination of values is given by $E_{tot}^{e^+e^-} = 
3.87 \times 10^{52}$ erg and $B = 1.5 \times 10^{-4}$, but this would imply a very high value for the 
Lorentz factor at the transparency of $\sim$ 5000. In turn, this value would imply a thermal spectrum of the
P-GRB peaking at around $300$ keV, which is in contrast with the observed temperature of $kT_{1st} = 58$ keV. 

Anyway, if we consider all the 50s emission as the extended afterglow, and assuming a ''virtual'' P-GRB lasting 10s and
below the nominal Fermi detector threshold, we conclude that this P-GRB should have been
detected by Fermi (Ruffini et al., Adv. Space Res. submitted). We can then 
conclude that in no way we can interpret this episode either as a
P-GRB of the second episode or, as proved here, as a separate GRB.

\section{A DIFFERENT EMISSION PROCESS}

\begin{table*}
\centering
\caption{Time-resolved spectral analysis of the first episode in GRB 090618. We have considered seven time intervals, as described in the text, and we used two spectral models, whose best-fit parameters are shown here. } % title ofable
\label{tab:2} % is used to refer this table in the text
{
\begin{tabular}{l c c c c c c c}
\hline\hline
Time  & $\alpha$ & $\beta$ & $E_0$ (keV) & $\tilde{\chi}^2_{BAND}$ & $kT$ (keV) & $\gamma$ & $\tilde{\chi}^2_{BB+po}$ \\ % table heading
\hline % inserts single horizontal line 
0 - 5 & -0.45 $\pm$ 0.11 & -2.89 $\pm$ 0.78 & 208.9 $\pm$ 36.13 & 0.93 & 59.86 $\pm$ 2.72 & 1.62 $\pm$ 0.07 & 1.07  \\
5 - 10 & -0.16 $\pm$ 0.17 & -2.34 $\pm$ 0.18 & 89.84 $\pm$ 17.69 & 1.14 & 37.57 $\pm$ 1.76 & 1.56 $\pm$ 0.05 & 1.36 \\
10 - 17 & -0.74 $\pm$ 0.08 & -3.36 $\pm$ 1.34 & 149.7 $\pm$ 21.1 & 0.98 & 34.90 $\pm$ 1.63 & 1.72 $\pm$ 0.05 & 1.20 \\
17 - 23 & -0.51 $\pm$ 0.17 & -2.56 $\pm$ 0.26 & 75.57 $\pm$ 16.35 & 1.11 & 25.47 $\pm$ 1.38 & 1.75 $\pm$ 0.06 & 1.19 \\
23 - 31 & -0.93 $\pm$ 0.13 & unconstr. & 104.7 $\pm$ 21.29 & 1.08 & 23.75 $\pm$ 1.68 & 1.93 $\pm$ 0.10 & 1.13 \\           
31 - 39 & -1.27 $\pm$ 0.28 & -3.20 $\pm$ 1.00 & 113.28 $\pm$ 64.7 & 1.17 & 18.44 $\pm$ 1.46 & 2.77 $\pm$ 0.83 & 1.10 \\
39 - 49 & -3.62 $\pm$ 1.00 & -2.19 $\pm$ 0.17 & 57.48 $\pm$ 50.0 & 1.15 & 14.03 $\pm$ 2.35 & 3.20 $\pm$ 1.38 & 1.10 \\
\hline
\end{tabular}
}
\end{table*}

A detailed spectral analysis of the first episode showed a strong spectral evolution, see Table \ref{tab:2}, 
where we have taken into account two different models, a Band model and a blackbody plus an extra power-law 
component, in order to explain this first emission episode. We have seen, in the case of the black body plus 
power-law, that the temperature varies from the initial value of 54 keV to 14 keV, where the extra power-law
 non-thermal component is assumed to be related to the blackbody by some process which will be studied in 
the future. Assuming a non-relativistic expansion for this first episode, we computed the evolution of the 
blackbody radius just from the luminosity observed:
\begin{equation}\label{eq:radius}
r_{em} = \frac{[\phi_{obs}/(\sigma T^4_{obs})]^{1/2} D}{(1+z)^2}.
\end{equation}
, where $\phi_{obs}$ is the observed thermal flux, $\sigma$ the Stefan constant and $D$ the luminosity distance.
We have also obtained a lower limit estimate of the mass of the proto-black hole by considering the sole blackbody energy emitted in the first episode, given by the thermal luminosity times the emission time, and its origin as due to the gravitational energy:
\begin{equation}\label{eq:virial2}
E_{iso} = 4 \pi r_{em}^2 \sigma T^4 \Delta t = \frac{3}{5} \frac{G M^2_{\rm pbh}}{r_{em}}
\end{equation}
where $E_{iso}$ is the total isotropic energy of the event, $\Delta t$ is the time duration of the first episode in the rest frame and $M_{\rm pbh}$ denotes the proto-black hole mass. 
We then obtained 
\begin{equation}\label{eq:Mpbhupper}
M_{\rm pbh} \gtrsim \sqrt{\frac{20 \pi r_{em}^3 \sigma T^4 \Delta t}{3 G}} \sim 4 M_\odot.
\end{equation}
as a lower limit for proto-black hole mass. The physical explanation for the proto-black hole emission is to be attributed to the processes of shells' mixing that happen in the last stages of the final collapse of a very massive star \cite{8}.  
Full details on the observational and theoretical evidences of the double episodes nature of GRB 090618 can be found in Izzo et al., A\&A submitted.

\bigskip % extra skip inserted

\begin{thebibliography}{9}   % Use for  1-9  references
%\begin{thebibliography}{99} % Use for 10-99 references

%\bibitem{accelconf-ref}
%http://www.cern.ch/accelconf

\bibitem[Ruffini et al.(2009)]{1}
Ruffini et al., AIPC, 1132, 199 (2009)

\bibitem[Gehrels et al.(2004), Burrows et al.(2005)]{2}
Gehrels, N. et al., ApJ, 611, 1005 (2004); Burrows, S. D., et al., Space Sci. Rev., 120, 16 (2005)

\bibitem[Meegan et al.(2009)]{3}
Meegan, C., Lichti, G., Bhat, P. N., et al., ApJ, 702, 791 (2009)

\bibitem[Rao et al.(2011)]{4}
Rao, A. R., Malkar, J. P., Hingar, M. K., et al., ApJ, 728, 42 (2011)

\bibitem{5}  
http://heasarc.gsfc.nasa.gov/lheasoft

\bibitem[Ruffini et al.(2000)]{6}
Ruffini, R., Salmonson, J.D., Wilson, J.R. And Xue, S.S., A\&A, 359, 855 (2000)

\bibitem[Arnett \& Meakin(2011)]{8}
Arnett, W. D., and Meakin,C., ApJ, 733, 78 (2011)


%\bibitem{templates-ref}
%http://www.cern.ch/accelconf/templates.html

\end{thebibliography}

\end{document}